\documentclass[aps,pra,floatfix]{revtex4}

\usepackage{mathrsfs}
\usepackage{float}
\usepackage{graphicx}
\usepackage[english]{babel}
\usepackage{amsmath}
\usepackage{amssymb}
\usepackage{extarrows}
\usepackage{color}
\usepackage[colorlinks=true, letterpaper=true, pdfstartview=FitV, linkcolor=blue, citecolor=blue, urlcolor=blue]{hyperref}
\usepackage{CJK}
\usepackage{slashed}
\usepackage{bm}
\usepackage{cancel}
\usepackage{txfonts}

\newcommand{\me}{\mathrm{e}}
\newcommand{\mi}{\mathrm{i}}

\newcommand{\dif}{\mathrm{d}}

\newcommand{\bk}{\mathbf{k}}

\newcommand{\bx}{\mathbf{x}}

\allowdisplaybreaks
\begin{document}
\title{Thermostatistics of a $q$-deformed Relativistic Ideal Fermi Gas}
\author{ Xu-Yang Hou$^{1}$,H. Yan$^{2}$, Hao Guo$^{1\ast}$}
\affiliation{$^1$Department of Physics, Southeast University, Nanjing 211189, China}
\affiliation{$^2$Key Laboratory of Neutron Physics, Institute of Nuclear Physics and Chemistry, CAEP, Mianyang 621900, Sichuan, China}

\begin{abstract}
In this paper, we formulate a $q$-deformed many-body theory for the relativistic Fermi gas and discuss the effects of the deformation parameter $q$ on  physical properties of such systems. Since antiparticle excitations appear in the relativistic regime, a suitable treatment to the choice of deformation parameters for both fermions and antifermions must be carefully taken in order to get a consistent theory.
By applying this formulation, we further study the thermostatistic properties of a $q$-deformed ideal relativistic Fermi gas. It can be shown that even in the noninteracting scenario, the system exhibits interesting characteristics which are significantly different from ordinary Fermi gases. Explicitly, antiparticles may become dominant due to the shift of chemical potential by the deformation parameter $q$.
This may build a solid foundation for the further studies of $q$-deformed relativistic interacting systems. We also apply our model to study the electron gas in a white dwarf. The effect of the deformation parameter on Chandrasekhar limit is discussed.
\end{abstract}
\email{guohao.ph@seu.edu.cn}
\maketitle

\section{Introduction}\label{Sec.1}
About four decades ago, the studies on exactly solvable models inspired the concepts of quantum groups and the associated algebras\cite{Drinfeld,JimboLMP86,JimboBook,qo0,Vega}.
The quantum group, for example SU(2)$_q$, can be realized by the $q$-analogue of the SU(2) algebra\cite{qaJPA89,qaJPA89b}. The Jordan-Schwinger mapping of this quantum algebra is the $q$-analogue of harmonic oscillators, which satisfy a $q$-deformed commutative relation\cite{qaJMP76}, called $q$-deformed algebra.
The $q$-deformed algebra has obtained a lot of study interests and has become the subject of intensive research works due to its beautiful mathematical structure and rich physical significance.
It has found many applications in various topics including the string theory\cite{qstring2D}, quantum gravity\cite{qaqg92}, dS/CFT and AdS/CFT correspondance\cite{qaPRD04,qaPTEP16}, quantum optics\cite{qo1,qo2,qo3,qo4,qo5}, many-body systems\cite{ISI:000263121100001} and the nuclear physics\cite{qfnp,qbcs3,qbcs1,qbcs2,qnjl,qnclp01,qnclp03,qnclp03a,qnclp11}.
 Importantly, people have begun to apply it to study more realistic phenomena, such as the black body radiation\cite{comqJPA94}, dark energy\cite{qaIJMPD17}, Bose-Einstein condensation\cite{qaBEC02}, the emissivity of the light fermionic dark matter in the cooling of the supernova SN1987A\cite{qsn17}, and thermal properties of solid\cite{qaSP12,qaSP18}.

Deformed many-particle systems are formed by indistinguishable particles satisfying the $q$-deformed commutative or anti-commutative relations, which form representations of $q$-deformed algebras. They are essentially different from the well-known anyons since the former can exist in arbitrary dimensions while the latter is only meaningful in two-dimensional systems. In some literatures, the $q$-deformed particles are called $q$-particles or simply quons. There have been intensive studies on the ideal many-quon systems\cite{VPJC1,qh11,qfs1,qfs2,qfs3,qfs4,Algin11,QF1,QF2,qfm04,qaPRE12}. In our recent paper\cite{OurJStat18}, we successfully constructed a $q$-deformed many-body theory, the $q$-deformed BCS theory, for the most famous interacting fermionic system, the superconductor. By applying it to the $q$-deformed interacting Fermi gases, we obtained some interesting physical predictions\cite{HXpaper}.

Until now, most research works focus on the nonrelativistic systems or low energy systems in which no antiparticle excitations onset. It is natural to generalize the current studies to the high energetic/relativistic regime. In this paper, we try to construct a many-body theory for the ideal $q$-deformed relativistic Fermi gas. Future studies on interacting $q$-deformed relativistic Fermi gases must be built based on this foundation. Note both particle and antiparticle excitations have their deformation parameters, we must carefully tune the parameters to get a self-consistent theory. It will be found that even in this noninteracting situation, the $q$-deformed relativistic Fermi gas already possesses many interesting features due to the deformation parameter $q$.



The rest of the paper is organized as follows. In Sec.\ref{Sec.2}, we give a self-consistent deformed relation for operators of $q$-fermions/$q$-antifermions, and discuss its basic properties. In Sec.\ref{Sec.2a}, we further construct a finite temperature many-body theory of $q$-deformed relativistic ideal Fermi gas based on the Green's function formalism. In Sec.\ref{Sec.3}, we derive the associated equations of states and discuss the thermodynamics. In Sec.\ref{Sec.4}, we give numerical analysis to give a deep understanding of the characteristics of such systems.
In Sec.\ref{Sec.4b}, we apply our model to study the electron gas inside a white dwarf. The effect of the deformation parameter on Chandrasekhar limit is studied.
The conclusion is summarized in the end.



\section{$q$-deformed Algebra}\label{Sec.2}
Throughout this paper, we adopt the natural unit system with $\hbar=c=k_B=1$ to simplify the notations.
The deformation parameter $q$ is taken as a positive real number. There are also studies on complex deformation parameter\cite{comqJPA94}.
In a relativistic $q$-deformed many-fermion system, there exist both fermionic and antifermionic excitations. Before starting to formulate a $q$-deformed algebra for these $q$-particles, we need to deal with an important question: how to assign the deformation parameters to $q$-fermions and $q$-antifermions respectively? Or more precisely, what is the relation between these two deformation parameters? It has been shown in \cite{VPJC2,OurJStat18} that in a $q$-deformed many-particle system, the chemical potential is shifted by the deformation parameter. It is known that the chemical potential $\mu$ is defined by the energy change as one extra particle is added to a quantum system containing a large number of particles. Moreover, in a relativistic system, adding an antiparticle is equivalent to removing a particle. Thus, the chemical potential of the corresponding antiparticle is $-\mu$. For $q$-deformed many-fermion systems, the chemical potential $\mu$ of a $q$-fermion is shifted to $\mu+T\ln q$ where $T$ is the temperature\cite{OurJStat18}. Thus, it is reasonable to infer that the chemical potential of a $q$-antifermion is shifted by the deformation parameter as
\begin{align}\label{a0}
-\mu-T\ln  q=-\mu+T\ln q^{-1}.
\end{align}
That is to say, the deformation parameter for a deformed antifermion is $q^{-1}$ if the deformation parameter of the deformed fermion is $q$.
Based on this analysis, the deformed canonical (anti)commutation relations can be constructed as follows.
We introduce the field operator $a_\bk$ to denote annihilation operator for the $q$-fermion with momentum $\bk$, and $b^\dagger_\bk$ the creation operator for the $q$-antifermion with momentum $\bk$. We assume both types of operators satisfy the Viswanathan-Parthasarathy-Jagannathan-Chaichian (VPJC) algebra\cite{Algin11}, i.e.,
\begin{align}\label{a1}
&a_{\bk}a^\dagger_{\bk'}+qa^\dagger_{\bk'}a_{\bk}=(2\pi)^3\delta_{\bk\bk'},\quad
\left[\hat{n}^-_{\bk}, a_{\bk'}\right]=-a_{\bk}\delta_{\bk\bk'},\quad
\left[\hat{n}^-_{\bk}, a^\dagger_{\bk'}\right]=a^\dagger_{\bk}\delta_{\bk\bk'},\notag\\
&qb_{\bk}b^\dagger_{\bk'}+b^\dagger_{\bk'}b_{\bk}=(2\pi)^3\delta_{\bk\bk'},\quad
\left[\hat{n}^+_{\bk}, b_{\bk'}\right]=-b_{\bk}\delta_{\bk\bk'},\quad
\left[\hat{n}^+_{\bk}, b^\dagger_{\bk'}\right]=b^\dagger_{\bk}\delta_{\bk\bk'},\notag\\
&a_{\bk}b_{\bk'}+qb_{\bk'}a_{\bk}=0,\quad b^\dagger_{\bk}a^\dagger_{\bk'}+qa^\dagger_{\bk'}b^\dagger_{\bk}=0, 
\end{align}
and all other anti-commutations vanish. Here $\hat{n}^\mp_\bk$ is the number operator for $q$-deformed fermions/antifermions, and the factor $(2\pi)^3$ is included for the convenience of normalization. The operators $[\hat{n}^-_\bk]\equiv \frac{1}{(2\pi)^3}a^\dagger_{\bk}a_{\bk}$ is called the deformed number operator for $q$-fermions \cite{VPJC1,VPJC2,Algin11}, and satisfies $[\hat{n}^-_\bk]|n_\bk\rangle_-=[n_\bk]_-|n_\bk\rangle_-$ where $|n_\bk\rangle_-$ with $n_\bk\ge0$ forms the Fock space of $q$-fermions. By applying the first identity of Eq.(\ref{a1}), we get a recurrence equation (Details of derivations can also be found in Appendix.\ref{app0})
\begin{align}\label{aa30}
[n_\bk+1]_-=1-q[n_\bk]_-.
\end{align}
The solution of Eq.(\ref{aa30}), which is called the $q$-fermionic basic number, is given by
\begin{align}\label{aamp}
[n_\bk]_-=\frac{1-(-1)^{n_{\bk}}q^{n_{\bk}}}{1+q}.
\end{align}
The corresponding state vector of the Fock space for $q$-fermions is
\begin{align}\label{aa50}
|n_\bk\rangle_-=\frac{1}{\sqrt{(2\pi)^{3n_{\bk}}}}\frac{(a^\dagger_\bk)^{n_\bk}}{\sqrt{[n_\bk]_-!}}|0\rangle
\end{align}
where $[n_\bk]_-!=[n_\bk]_-[n_\bk-1]_-\cdots [2]_-[1]_-$.
When $q=1$, the Pauli exclusion principle is recovered from Eq.(\ref{aamp}) since $\lim_{q\rightarrow 1}[2]_-=0$. 
Moreover, Eq.(\ref{aamp}) shows $[n]_-$ is negative if $n$ is an even positive integer if $q>1$. Thus, we require $0<q\le 1$ to ensure that $|n_\bk\rangle_-$ has a positive norm, as pointed out in Ref.\cite{VPJC1}.

We are interested in whether these discussions can be generalized to $q$-antifermions. If we introduce deformed number operator $[\hat{n}^+_\bk]\equiv \frac{1}{(2\pi)^3}b^\dagger_{\bk}b_{\bk}$ for $q$-fermions/$q$-antifermions, and assume $[\hat{n}^+_\bk]|n_\bk\rangle_+=[n_\bk]_+|n_\bk\rangle_+$ where $|n_\bk\rangle_+$ with $n_\bk\ge0$ forms the Fock space of $q$-antifermions, by applying the second line of Eq.(\ref{a1}), similar analysis shows
\begin{align}\label{ba30}
q[n_\bk+1]_+=1-[n_\bk]_+=1-q^{-1}(q[n_\bk]_+).
\end{align}
Comparing Eq.(\ref{ba30}) with Eq.(\ref{aa30}), the basic fermionic number for $q$-antifermions can be inferred from Eq.(\ref{aamp})
\begin{align}\label{aamp2}
 [n_\bk]_+=q^{-1}\frac{1-(-1)^{n_{\bk}}q^{-n_{\bk}}}{1+q^{-1}}=\frac{1-(-1)^{n_{\bk}}q^{-n_{\bk}}}{1+q}.
\end{align}
However, when $0<q<1$, Eq.(\ref{aamp2}) indicates that $[n_\bk]_+<0$ if $n_\bk$ is an even positive integer. That means the state vector \begin{align}\label{ba50}|n_\bk\rangle_+=\frac{1}{\sqrt{(2\pi)^{3n_{\bk}}}}\frac{(b^\dagger_\bk)^{n_\bk}}{\sqrt{[n_\bk]_+!}}|0\rangle\end{align} has a negative norm when $n_\bk$ is both positive and even.
These states, which may be referred to as the ghost states, often appear in the gauge field theory\cite{Peskin_book}, and also in the models involved with higher derivatives\cite{IJTP95}.
To understand where this puzzle comes, we examine how the Fock space of $q$-antifermions is built.
Note an implicit premise, which requires that the ground state $|0\rangle$ is annihilated by $a_\bk$ and $b_\bk$ simultaneously, is needed for the construction of Fock states (\ref{aa50}) and (\ref{ba50}). This means that $|0\rangle$ is an ``empty'' state with nothing, which is a common assumption in quantum field theory\cite{Peskin_book} to ensure that all excitations of $|0\rangle$ have positive energy. Moreover, this condition is introduced when quantizing the Hamiltonian of a fermionic field theory. However, we haven't started to build the field theory for a many-$q$-fermion system yet, and no Hamiltonian or Lagrangian is introduced until now.
 In the rest part of this paper, we will restrict the values of $q$ as $0<q\le 1$ to ensure that all Fock states in particle sector has nonnegative norm.
 Since the algebra (\ref{a1}) is crucial to construct a $q$-deformed relativistic many-body theory, we must carefully check that the states $(b^\dagger_\bk)^n|0\rangle$ with $n\ge2$ will not appear in our model.
Obviously, this is easily achieved in the field theory for ordinary fermionic systems since more than one fermions are not allowed to occupy a same state. In the later chapter, we will show that the states $(b^\dagger_\bk)^n|0\rangle$ with $n\ge2$ does not appear when evaluating the partition function in our model.

\section{Many-body Theory}\label{Sec.2a}

To introduce the Hamiltonian of a noninteracting relativistic $q$-Fermi-gas, it must be noticed that the deformed number operator $[\hat{n}^+_\bk]$ can not be included since previous discussions indicate that it brings in states of negative norm.
Typically, the Hamiltonian is obtained by generalizing that of a nonrelativistic $q$-gas\cite{Algin11,VPJC1,VPJC2} by including a term for the $q$-antifermion sector
\begin{equation}\label{H0}
H=\sum_{\mathbf{k}}\left[(\epsilon_{\mathbf{k}}-\mu)\hat{n}^-_\bk+(\epsilon_{\mathbf{k}}+\mu)\hat{n}^+_\bk\right],
\end{equation}
where $m$ is the mass and $\epsilon_{\mathbf{k}}=\sqrt{\mathbf{k}^2+m^2}$. $\epsilon_{\mathbf{k}}\mp\mu$
is the energy dispersion of the $q$-deformed fermion/antifermion.
To formulate the many-body theory of a $q$-gas, we introduce the $q$-fermion field $\psi(\mathbf{x})$ in the coordinate space, which is given by Eq.(\ref{psi}). We also introduce the 4-momentum $k^\mu=(\epsilon_\mathbf{k},\mathbf{k})$, $\sigma^{\mu}=(1,\vec{\sigma})$ and $\bar{\sigma}^{\mu}=(1,-\vec{\sigma})$ where $\vec{\sigma}=(\sigma^1,\sigma^2,\sigma^3)^T$ are the pauli matrices to express our formulation in a compact form.
In the imaginary-time formalism of finite temperature field theory, we define the finite temperature Heisenberg operator
$\psi(x)=\me^{H\tau}\psi(\mathbf{x})\me^{-H\tau}$ where $x=(\tau,\mathbf{x})$ with $\tau=\mi t$ being the imaginary time.
The field $\psi(x)$ contains both fermionic and anti-fermionic excitations, which can be expanded in the momentum space as
\begin{eqnarray}\label{p1}
\psi (x)=\sum_{\mathbf{k}}\frac{1}{\sqrt{\epsilon_{\mathbf{k}}}}\Big\{
\left[ \begin{array}{ccc}
\sqrt{k\cdot\sigma}\eta^{L}_{\mathbf{k}} \\
\sqrt{k\cdot\bar{\sigma}}\eta^{L}_{\mathbf{k}}
\end{array}\right]a_\mathbf{k}\me^{-\mi k_-\cdot x}+
\left[ \begin{array}{ccc}
\sqrt{k\cdot\sigma}\eta^{R}_{\mathbf{k}} \\
-\sqrt{k\cdot\bar{\sigma}}\eta^{R}_{\mathbf{k}}
\end{array}\right]b^{\dagger}_\mathbf{k}\me^{\mi k_+\cdot x}\Big\}, 
\end{eqnarray}
where $k_\mp^\mu=(\epsilon_\mathbf{k}\mp\mu,\mathbf{k})$, $\me^{\mp\mi k_\mp\cdot x}=\me^{\pm t(\epsilon_\mathbf{k}\mp\mu)+\mi (\pm\mathbf{k}\cdot\mathbf{x}+\frac{\pi}{2})}$,
$\eta^{L}_{\mathbf{k}}$ and $\eta^{R}_{\mathbf{k}}$ are the two eigenvectors of $\vec{\sigma}\cdot\hat{\mathbf{k}}$ with different handedness, and the scalar product between two 4-vectors are given in the Appendix.\ref{app1}. Explicitly, $\vec{\sigma}\cdot\mathbf{k}\eta^{R,L}_{\mathbf{k}} =\pm|\mathbf{k}|\eta^{R,L}_{\mathbf{k}}$ with
\begin{equation}
\eta^{L}_{\mathbf{k}}=
\left[ \begin{array}{ccc}
-\mbox{sin}\frac{\theta_{\mathbf{k}}}{2}\me^{-\mi\phi_{\mathbf{k}}} \\
\mbox{cos}\frac{\theta_{\mathbf{k}}}{2}
\end{array}\right] \qquad  \eta^{R}_{\mathbf{k}}=
\left[ \begin{array}{ccc}
\mbox{cos}\frac{\theta_{\mathbf{k}}}{2} \\
\mbox{sin}\frac{\theta_{\mathbf{k}}}{2}\me^{\mi\phi_{\mathbf{k}}}
\end{array}\right],
\end{equation}
 where $\theta_{\mathbf{k}}$ and $\phi_{\mathbf{k}}$ are the polar and azimuthal angels of vector $\mathbf{k}$. It can be shown that the fermion field (for simplicity, $\psi(\mathbf{x})$ and $\psi(x)$ are still called fermion field hereafter) satisfies the following $q$-deformed anti-commutative relation (for details, please refer to Appendix.\ref{app2})
\begin{align}\label{qacr}
\psi_a (\mathbf{x})\psi_b^\dagger (\mathbf{x}')+q\psi_b^\dagger (\mathbf{x}')\psi_a (\mathbf{x})=\delta_{ab}\delta(\mathbf{x}-\mathbf{x}').
\end{align}
The equation of motion of the fermion field can be obtained by the Heisenberg equation $\frac{\partial \psi(x)}{\partial \tau}=[H, \psi(x)]$. Applying the algebras (\ref{a1}), and plugging in the expressions (\ref{H0}) and (\ref{p1}), we have
\begin{equation}\label{eom}
\gamma^0\frac{\partial \psi(x)}{\partial \tau}=(\mi\vec{\gamma}\cdot\nabla-m+\mu\gamma^0) \psi(x),
\end{equation}
where $\gamma^0$ and $\vec{\gamma}=(\gamma^1,\gamma^2,\gamma^3)^T$ are gamma matrices.
Details can be found in the Appendix.\ref{app3}. A Lagrangian of which the Eular's equation of motion is exactly Eq.(\ref{eom}) is given by
\begin{equation}
L=\int d^3\mathbf{x}\bar{\psi}(\mi\gamma^{\mu}\partial_{\mu}-m+\mu\gamma^{0})\psi,
\end{equation}
where $\bar{\psi}\equiv \psi^\dagger\gamma^0$.
The deformed Green's function of the fermion field is defined as
\begin{align}
G_{ab}(x,x^{\prime})=-\langle T_{\tau}[\psi_a(x)\bar{\psi}_b(x^{\prime})]\rangle=-\langle \psi_a(x)\bar{\psi}_b(x^{\prime})\rangle\theta(\tau-\tau') +q\langle \bar{\psi}_b(x')\psi_a(x)\rangle\theta(\tau'-\tau).
\end{align}
It is reasonable to assume that the deformed Green's function has a translational symmetry in spacetime, i.e. $G(x,x')=G(x-x')\equiv G(\tau-\tau',\mathbf{x}-\mathbf{x}')$. Moreover, It can be shown that the deformed Green's function has a different periodic property from that of the ordinary fermionic Green's function\cite{OurJStat18}:
\begin{align}\label{p3}
G(-\beta<\tau-\tau'<\beta,\bx-\bx')=-qG(\tau-\tau'+\beta,\bx-\bx'),
\end{align}
where $\beta=1/T$ is the inverse temperature. This leads to the fact that the fermionic Matsubara
frequency gets an extra imaginary part, which can be absorbed into the chemical potential. The details will be shown later.
By using the equations of motion (\ref{eom}), we have
\begin{eqnarray}
[\gamma^{0}\partial_{\tau}G(x,x^{\prime})]_{ab}&=&-\delta(\tau-\tau^{\prime})\gamma^{0}_{ac}\langle\left[\psi_{  c}(x)\psi^\dagger_{  d}(x^{\prime})+q\psi^\dagger_{  d}(x^{\prime})\psi_{  c}(x)\right]\rangle\gamma^{0}_{db}-\langle T_{\tau}(\gamma^{0}\partial_{\tau}\psi(x)\bar{\psi}_{ }(x^{\prime}))\rangle_{ab} \nonumber \\
&=&-\delta_{ab}\delta(\tau-\tau^{\prime})\delta(\mathbf{x}-\mathbf{x}^{\prime})-[(\mi\vec{\gamma}\cdot\nabla-m+\mu\gamma^{0})\langle T_{\tau}(\psi_{ }(x)\bar{\psi}(x^{\prime}))\rangle]_{ab} .
\end{eqnarray}
Rearrange the equation we get
\begin{equation}\label{eomp}
[-\gamma^{0}\partial_{\tau}+\mi\vec{\gamma}\cdot\nabla-m+\mu\gamma^{0})]G(x,x^{\prime})=\delta(x-x^{\prime})\mathbf{1}_{4\times4}
\end{equation}
which is the equation of motion for the deformed Green's function.
Using the property (\ref{p3}) and solving Eq.(\ref{eomp}) in the momentum space, the solution to the deformed Green's function is
\begin{equation}\label{G-1}
G^{-1}(K)=\left(\mi\omega_{n}+\frac{\ln q}{\beta}+\mu\right)\gamma^{0}-\vec{\gamma}\cdot\mathbf{k}-m,
\end{equation}
where
$K=(\mi\omega_{n},\mathbf{k})$ is the fermionic 4-momentum at finite temperature with $\omega_{n}$ being the ordinary fermionic Matsubara frequency $\omega_{n}=(2n+1)\pi T$ $(n=0, \pm1, \pm2, \dots)$.
As we have pointed out, the Matsubara frequency obtains an imaginary part $\frac{1}{\beta}\ln q= T\ln q$. Thus, the chemical potential $\mu$ is shifted to $\mu+T\ln q$, which agrees with our former discussions at the beginning of the section.
Introducing the energy projectors for particle and antiparticle sectors
\begin{equation}
\Lambda_{\pm}(\mathbf{k})=\frac{1}{2}\left(1\pm\frac{\gamma^{0}(\vec{\gamma}\cdot\mathbf{k}+m)}{\epsilon_{\mathbf{k}}}\right)
\end{equation}
and taking the inverse of Eq.(\ref{G-1}), the Green's function can be expressed as (see Appendix.\ref{app4})
\begin{align}\label{GK}
G(K)
&=\left[\frac{\Lambda_{+}(\mathbf{k})}{\mi\omega_{n}-\left(\xi^{-}_{\mathbf{k}}-\frac{\ln q}{\beta}\right)}+\frac{\Lambda_{-}(\mathbf{k})}{\mi\omega_{n}+\left(\xi^{+}_{\mathbf{k}}+\frac{\ln q}{\beta}\right)}\right]\gamma^{0},
\end{align}
where $\xi^{\pm}_{\mathbf{k}}=\epsilon_{\mathbf{k}}\pm \mu$.
After taking complex continuation as $\mi\omega_n\rightarrow \omega+\mi 0^+$, the two poles of the Green's function indicate that the energy dispersions of  quasi $q$-fermions/quasi $q$-antifermions (which will be simplified as quasi-fermions/quasi-antifermions hereafter) are $\pm(\xi^\mp\mp T\ln q)$ respectively. However, the energy spectrum of a physical system can not be negative. Since adding an antiparticle is equivalent to removing a particle, the energy dispersion of the quasi $q$-antifermion is, in fact, $\xi^++T\ln q$.
Thus, the chemical potential of quasi-fermions/quasi-antifermions are shifted as $\pm (\mu+T\ln q)$ respectively, which indeed coincides with Eq.(\ref{a0}) and the related discussions. Hence, our choice of the deformation parameters is self-consistent.

In the nonrelativistic limit $|\mathbf{k}|\ll m$, the energy dispersion of the quasi-fermion is $\xi^{-}_{\mathbf{k}}\simeq \frac{\mathbf{k}^{2}}{2m}-(\mu-m)$, so the nonrelativistic result is recovered and the quantity $\mu-m$ plays the role of $\mu$.
Moreover, since $\epsilon_\bk\simeq m$ in the nonrelativistic limit, then $\Lambda_{+}(\mathbf{k})\simeq 1$, $\Lambda_{-}(\mathbf{k})\simeq 0$. Hence the main contribution to the Green's function comes from the particle sector
\begin{align}
G(K)\simeq\frac{\gamma^0}{\mi\omega_{n}+\frac{\ln q}{\beta}-\xi^{-}_{\mathbf{k}}},
\end{align}
which essentially reproduces the non-relativistic Green's function\cite{Fetter_book}.

\section{Equation of States and Thermodynamics}\label{Sec.3}
The most important equation of states is the particle number equation. It is obtained by the ensemble average of the deformed fermion number operator $\langle \psi^+\psi\rangle$, which can be further evaluated by the Green's function. By using the property $\text{Tr}(\Lambda_{\pm}(\mathbf{k}))=2$ and Eq.(\ref{GK}), we have
 \begin{eqnarray}\label{N1}
    N
    &=&\frac{1}{2}\sum_a\int d^3\bx\langle \psi^\dagger_a(\bx)\psi_a(\bx)\rangle=\frac{1}{2}\sum_a\int d^3\bx\langle \bar{\psi}_b(\bx)\gamma^0_{ba}\psi_a(\bx)\rangle=\frac{1}{2q}\int d^3\bx\textrm{Tr}[ G(x,x^+)\gamma^0]\nonumber\\
    &=&\frac{V}{2q}T\sum_{i\omega_n}\sum_{\bk}\textrm{Tr}[G(\mi\omega_n,\bk)\gamma^0]=\frac{V}{q}\sum_\bk\left[ f(\xi^-_\bk-T\ln q)+f(-\xi^+_\bk-T \ln q)\right],
    \end{eqnarray}
    where $V$ is the volume, $f(x)=1/(\me^{\beta x}+1)$ is the ordinary Fermi distributive function, and the factor of $1/2$ is included to remove the extra degrees of freedom. The particle number density $n\equiv\frac{N}{V}$ is inferred from Eq.(\ref{N1})
    \begin{align}\label{n0}
    n&=\frac{1}{q}\sum_\bk\left(\frac{1}{\me^{\beta(\epsilon_\bk-\mu-T\ln q)}+1}+\frac{1}{\me^{-\beta(\epsilon_\bk+\mu+T\ln q)}+1}\right)\notag\\&=\sum_\bk\left(\frac{1}{\me^{\beta\xi^-_\bk}+q}+\frac{1}{\me^{-\beta\xi^+_\bk}+q}\right)\rightarrow\sum_\bk\left(\frac{1}{\me^{\beta\xi^-_\bk}+q}-\frac{1}{q}\frac{1}{q\me^{\beta\xi^+_\bk}+1}\right),
    \end{align}
 where we have applied the fact $f(-x)=1-f(x)$, and omit the infinitely-large nonphysical quantity $\sum_\bk\frac{1}{q}$. The result is equal to the difference between the number densities of quasi-fermions and quasi-antifermions. That is, the former gives positive contribution while the latter gives negative contribution to the total particle number. This is consistent with the fact that they annihilate each other in a flash of energy when brought together. If $q=1$, our result reduces to the known result of ordinary relativistic Fermi gases~\cite{SEVILLA2017585,Reis20}.

      The particle number density can also be evaluated by taking derivative of the grand partition function $Z=\text{Tr}(\me^{-\beta H})$. There are some subtleties when calculating the trace. Previously, we have concluded that the energy dispersions of quasi-fermions/quasi-antifermions are $\xi^\mp\mp T\ln q$ respectively, in which the extra factor $T\ln q$ can be absorbed into the chemical potential as $\pm (\mu+T\ln q)$.
      Especially, the corrected chemical potential of quasi $q$-antifermions is $-\mu-T\ln q=-\mu+T\ln q^{-1}$, which confirms Eq.(\ref{a0}), i.e. the choice of deformation parameter $\frac{1}{q}$ is self-consistent.
      After doing this, the right-hand-side of the first line of Eq.(\ref{n0}) indicates that the quasi-fermions/quasi-antifermions can be equivalently thought of as satisfying the ordinary Fermi distribution instead of $q$-deformed Fermi distribution, except that the total number density is normalized by a factor of $\frac{1}{q}$. Thus, the equivalent Fock space of quasi-fermions/$q$-antifermions can be simply taken as spanned by $\{|0\rangle, |1\rangle\}$, just as the ordinary Fermi operators. In some sense, the Pauli exclusion principle still works here.
      Therefore, it is reasonable to evaluate the trace over this Fock space as
\begin{align}
Z
&=\sum_{\sum_\bk (n^-_\bk+n^+_\bk)=n;n^\pm_\bk=0,1}\prod_\mathbf{k}\otimes\left(\langle n^-_\bk |\me^{-\beta \sum_\bk(\xi^-_\bk-T\ln q) n^-_\bk}| n^-_\bk\rangle \langle n^+_\bk |\me^{-\beta \sum_\bk(\xi^+_\bk+T\ln q) n^+_\bk}| n^+_\bk\rangle\right)\notag\\
&=\prod_\mathbf{k}\left(\sum_{n^-_\bk=0,1}\me^{-\beta \sum_\bk(\xi^-_\bk-T\ln q) n^-_\bk}\sum_{n^+_\bk=0,1}\me^{-\beta \sum_\bk(\xi^+_\bk+T\ln q) n^+_\bk}\right)\notag\\
&=\prod_\mathbf{k}(1+qz\me^{-\beta\epsilon_\mathbf{k}})(1+q^{-1}z^{-1}\me^{-\beta\epsilon_\mathbf{k}}),
\end{align}
where $z=\me^{\beta\mu}$ is the fugacity. We emphasize here that the choice of this equivalent Fock space of quasi-antifermions indeed avoid including the negative-norm states $(b^\dagger_\bk)^n|0\rangle$ where $n\ge 2$ is an even integer. This answers the previous concern in Sec.\ref{Sec.2}. To check the validity of the partition function, we verify if it can give the reasonable form of equations of states.
It is straightforward to show that the number density is given by
 \begin{align}\label{n2}
n=\frac{1}{q}z\frac{\partial \ln Z}{\partial z},
 \end{align}
 where the inclusion of the deformation factor $\frac{1}{q}$ is consistent with our definition of the deformed Green's function. A brief proof of Eq.(\ref{n2}) is outlined in Appendix.\ref{app5}.
 The thermodynamical potential is defined as $\Omega=-\frac{1}{\beta}\ln Z$. Thus, Eq.(\ref{n2}) is equivalent to the well-known form (also in a deformed form)
 \begin{align}
n=-\frac{1}{q}\frac{\partial \Omega}{\partial \mu}.
 \end{align}
 The pressure is given by $PV=-\Omega$, which leads to another equation of state
 \begin{align}\label{ThP}
PV=T\ln Z=-\sum_\bk\left(\xi^-_\bk+\xi^+_\bk-T\ln\frac{1}{\me^{\beta\xi^-_\mathbf{k}}+q}-T\ln\frac{1}{\me^{\beta\xi^+_\mathbf{k}}+\frac{1}{q}}\right).
\end{align}
The entropy can be obtained by taking derivative of the partition function, or thermodynamical potential, as $S=-\frac{\partial \Omega}{\partial T}$. A straightforward calculation shows
 \begin{align}\label{S}
 S
&=-\sum_\bk\Big(f(\xi^-_\bk-T\ln q)\ln[\frac{1}{q}f(\xi^-_\bk-T\ln q)]+
f(-\xi^-_\bk+T\ln q)\ln\left[f(-\xi^-_\bk+T\ln q)\right]\notag\\
 &\qquad\quad +f(\xi^+_\bk+T\ln q)\ln\left[qf(\xi^+_\bk+T\ln q)\right]+f(-\xi^+_\bk-T\ln q)\ln\left[f(-\xi^+_\bk-T\ln q)\right]\Big).
\end{align}
The internal energy is defined as
 \begin{align}\label{E}
 E=-\frac{1}{q}\frac{\partial \ln Z}{\partial \beta}+\mu n= \frac{1}{q}(\Omega+TS)+\mu n.
  \end{align}
 Substitute Eqs.(\ref{n0}), (\ref{ThP}) and (\ref{S}), we get
 \begin{align}\label{Et}
 E=\sum_\bk\left(\frac{\epsilon_\bk}{\me^{\beta\xi^-_\bk}+q}+\frac{1}{q}\frac{\epsilon_\bk}{q\me^{\beta\xi^+_\bk}+1}\right).
 \end{align}
 Apparently, both particle and antiparticle modes give positive contributes to the total energy. The remaining thermodynamic functions are immediately found by $F=-P+\mu n$ and $G=\mu n$. All relations naturally reduce to the well-known results of ordinary Fermi gas when $q=1$.

 Now we give a qualitative study on the behaviors of the $q$-deformed relativistic ideal Fermi gas in the low temperature limit where the degenerate effect becomes dominant. Since $\epsilon_\bk=\mathbf{}\sqrt{\mathbf{k}^2+m^2}$, the summation over wavenumbers can be replaced as follows
\begin{equation}\label{Int}
\sum_\bk=\frac{gV}{(2\pi)^3}\int \dif^{3} \mathbf{k}=\frac{gV}{2\pi^{2}}\int\sqrt{\epsilon^{2}-m^{2}}\epsilon\dif \epsilon.
\end{equation}
where $g$ is the degeneracy of each single-particle momentum state and have been set to 2 in above discussion due to the spin degrees of freedom.
Applying this relation and taking integration by parts,
the equations of states, or the the thermodynamic potential, number density and internal energy given by Eqs.(\ref{ThP}), (\ref{n0}) and (\ref{E}) respectively become
\begin{align}\label{PV}
PV=-\Omega&
=\frac{gV}{6\pi^{2}}\int^{\infty}_{m}  \dif \epsilon \left[\frac{(\epsilon^{2}-m^{2})^{\frac{3}{2}} }{q^{-1}\me^{\beta(\epsilon-\mu)}+1}+\frac{ (\epsilon^{2}-m^{2})^{\frac{3}{2}} }{q\me^{\beta(\mu+\epsilon)}+1}\right],
\end{align}
\begin{equation}\label{NDE}
\frac{N}{V}=\frac{gV}{2q\pi^{2}}\int^{\infty}_{m}\dif \epsilon\left(\frac{\sqrt{\epsilon^{2}-m^{2}} \epsilon}{q^{-1}\me^{\beta(\epsilon-\mu)}+1}-\frac{\sqrt{\epsilon^{2}-m^{2}} \epsilon}{q\me^{\beta(\epsilon+\mu)}+1}\right),
\end{equation}
and
\begin{equation}\label{GSE}
E=\frac{gV}{2q\pi^{2}}\int^{\infty}_{m}\dif \epsilon\left(\frac{\sqrt{\epsilon^{2}-m^{2}}\epsilon^{2}}{q^{-1}\me^{\beta(\epsilon-\mu)}+1}+\frac{\sqrt{\epsilon^{2}-m^{2}}\epsilon^{2}}{q\me^{\beta(\epsilon+\mu)}+1}\right).
\end{equation}
Due to the relativistic energy dispersion, there is no way to get an explicit equation of state as $PV=\frac{2}{3}E$ in the case of nonrelativistic ideal Fermi gases.
In the zero temperature limit, the $q$-deformed Fermi distribution shown in Eq.(\ref{n0}) reduces to a step function
\begin{equation}\label{0TN}
\left(\frac{1}{\me^{\beta(\epsilon-\mu)}+q}-\frac{1}{q}\frac{1}{q\me^{\beta(\epsilon+\mu)}+1}\right) \overset{T\rightarrow 0}{\longrightarrow}
\left\{
\begin{array}{ccc}
0  &  & \epsilon > \mu \\
\frac{1}{q}  &  & \epsilon < \mu
\end{array}
\right\}
=\frac{1}{q}\theta(\mu-\epsilon).
\end{equation}
Here an implicit premise that $\epsilon+\mu>0$ at $T\rightarrow 0$ has been applied. Otherwise, only the second term of Eq.(\ref{n0}) gives non-zero contribution and the number density is accordingly negative. This means the high energetic quasi $q$-antifermion is dominant, which is impossible since all
kinetic degrees of freedom become frozen at $T\rightarrow 0$. Therefore, the zero temperature behavior of the $q$-deformed relativistic ideal Fermi gas is quite similar to the ordinary Fermi gas except the number density is normalized by $q$ and the energy dispersion is different. Thus, the ground state is also a Fermi sea (sphere), and there exists the Fermi level $E_F=\mu$ at $T=0$, which is the largest energy that a quasi $q$-fermion can possess. Using Eq.(\ref{0TN}), the number equation (\ref{NDE}) is readily evaluated as
\begin{align}\label{T0ND}
n=\frac{g}{6q\pi^{2}} (\mu^{2}-m^{2})^{\frac{3}{2}}.
\end{align}
This leads to the expression of Fermi level
\begin{equation}\label{FE}
E_{F}=\mu(T=0)=\sqrt{\left(\frac{6q\pi^{2}n}{g}\right)^{\frac{2}{3}}+m^{2}}=\sqrt{k_{F}^{2}+m^{2}},
\end{equation}
where
\begin{equation}\label{kf1}
k_{F}=\left(\frac{6q\pi^{2} n}{g} \right)^{\frac{1}{3}}=(3q\pi^2n)^{\frac{1}{3}}
\end{equation}
is the Fermi momentum. Note the shift $T\ln q$ of the chemical potential has no effect at $T=0$.
Similarly, by straightforward calculations, the thermodynamic potential and the internal energy are respectively given by
\begin{align}
PV&=\frac{gV}{6\pi^{2}}\left[\sqrt{\mu^{2}-m^2}\left(-\frac{5m^{2}}{8}\mu+\frac{\mu^{3}}{4}\right)+\frac{3}{8}m^{4}\ln\left(\frac{\mu}{m}+\sqrt{\frac{\mu^{2}}{m^{2}}-1}\right)    \right],\notag\\
\frac{E}{V}&=\frac{g}{2q\pi^{2}}\left[\sqrt{\mu^{2}-m^{2}}\left(-\frac{m^{2}}{8}\mu+\frac{\mu^{3}}{4}\right)-\frac{1}{8}m^{4}\ln\left(\frac{\mu}{m}+\sqrt{\frac{\mu^{2}}{m^{2}}-1}\right) \right] .
\end{align}
These are the zeroth order approximation of the equations of states at zero temperature limit.
It shows that a $q$-deformed relativistic ideal Fermi gas also exerts a finite pressure at zero temperature. This is because the structure of the ground state of the system is still a Fermi sea and the degenerate pressure dominates at $T=0$.

Next, we turn to the small but finite temperature, which gives the next order approximation to the equations of states. For simplicity, we focus on the thermodynamical potential, and other relations can be derived by following the same manner. Introducing the variable $x \equiv \beta(\epsilon-\mu-\frac{\ln q}{\beta})$, the first term on the right-hand-side of Eq.(\ref{PV}) yields
\begin{align}\label{FPVE}
I_{1}=\frac{1}{\beta^{4}}\int^{\infty}_{\beta m-\beta \mu-\ln q}  \dif x \frac{\big[(x+\beta\mu+\ln q)^{2}-(\beta m)^{2}\big]^{\frac{3}{2}} }{\me^{x}+1}.
\end{align}
Similarly, we define the variable $y \equiv \beta(\epsilon+\mu+\frac{\ln q}{\beta})$. Thus, the second term on the right-hand-side of Eq.(\ref{PV}) is rewritten as
\begin{align}\label{FPVE2}
I_{2}&=\frac{1}{\beta^{4}}\int^{\infty}_{\beta m+\beta \mu+\ln q}  \dif y \frac{\big[(y-\beta\mu-\ln q)^{2}-(\beta m)^{2}\big]^{\frac{3}{2}} }{\me^{y}+1}.
\end{align}
Note this integral is exponentially small in the limit $\beta m+\beta \mu+\ln q \rightarrow \infty$, then this term can be safely ignored.

To evaluate $I_1$, we introduce $a \equiv \beta\mu+\ln q$ and $b  \equiv\beta m$ to simplify the notation. Since $\mu$ is finite as $T\rightarrow 0$ and $m$ is a constant, then $a,b\rightarrow \infty$ if $T$ is small ($\beta\rightarrow \infty$). Hence the integral $I_1$ is further expressed as
\begin{equation}
I_{1}=\frac{1}{\beta^{4}}\int^{\infty}_{b-a}  \dif x \frac{\big[(x+a)^{2}-b^{2}\big]^{\frac{3}{2}} }{\me^{x}+1}.
\end{equation}
Now, change the variable $x\rightarrow -x$ in the first integral, and apply the identity $(\me^{-x}+1)^{-1}\equiv 1-(\me^{x}+1)^{-1}$, we get
\begin{align}\label{I12}
I_{1}&=\frac{1}{\beta^{4}}\int^{a-b}_{0}  \dif x \big[(a-x)^{2}-b^{2}\big]^{\frac{3}{2}}
+\frac{1}{\beta^{4}}\int^{\infty}_{0}  \dif x \frac{\big[(x+a)^{2}-b^{2}\big]^{\frac{3}{2}}-\big[(a-x)^{2}-b^{2}\big]^{\frac{3}{2}} }{\me^{x}+1} +\frac{1}{\beta^{4}}\int^{\infty}_{a-b}  \dif x \frac{\big[(a-x)^{2}-b^{2}\big]^{\frac{3}{2}}}{\me^{x}+1}.
\end{align}
The chemical potential $\mu$ is always larger than the mass $m$ at low enough temperatures, then $a-b$ is extremely large. Thus, the last term of Eq.(\ref{I12}) is exponentially small in this limit. Moreover, the numerator in the second integral can be approximated as
\begin{equation}
\big[(x+a)^{2}-b^{2}\big]^{\frac{3}{2}}-\big[(a-x)^{2}-b^{2}\big]^{\frac{3}{2}}\approx 6a(a^{2}-b^{2})^{\frac{1}{2}}x+\cdots, \qquad \text{when } a,b\rightarrow \infty.
\end{equation}
A straightforward calculation gives the asymptotic expansion to $I_1$
\begin{align}\label{I13}
I_{1}&=\left[\frac{1}{8}\left(\mu+\frac{\ln q}{\beta}\right)\sqrt{\left(\mu+\frac{\ln q}{\beta}\right)^{2}-m^{2}}\left[2\left(\mu+\frac{\ln q}{\beta}\right)^{2}-5m^{2}\right]+\frac{3}{8}m^{4}\ln\left(\frac{\beta\mu+\ln q}{\beta m}+\sqrt{\left(\frac{\beta\mu+\ln q}{\beta m}\right)^{2}-1}\right) \right] \notag \\
&+\frac{ \left(\mu+\frac{\ln q}{\beta}\right)\left[\left(\mu+\frac{\ln q}{\beta}\right)^{2}-m^{2}\right]\pi^{2}}{2\beta}+\cdots.
\end{align}
Therefore, the thermodynamic potential is
%
\begin{align}\label{{FPVE1}}
PV &\xlongequal[]{T\rightarrow 0}\frac{gV}{6\pi^{2}}I_{1}
\end{align}
where $I_1$ is given by Eq.(\ref{I13}).

\section{Numerical Analysis}\label{Sec.4}

To visualize more properties of the $q$-deformed relativistic ideal Fermi gas, we give a quantitative study based on numerical analysis. This can be accomplished by solving the number equation (\ref{n0}). For convenience, we fix the number density and solve $\mu$ at certain choices of $T$ and $q$. Eq.(\ref{kf1}) shows $n=\frac{k^3_F}{3q\pi^2}$ at $T=0$, i.e. $n$ depends on $q$. To give a better comparison between the results at different $q$'s, we choose the unit $k_F$ satisfying $n=\frac{k^3_F}{3\pi^2}$, which is the Fermi momentum of an un-deformed relativistic ideal Fermi gas. For simplicity, we still use the notation $k_F$ as our unit of momentum, which is different from Eq.(\ref{kf1}). The units of temperature and energy are chosen as $k_{B}T_{F}=E_{F}=\sqrt{k_{F}^{2}+m^{2}}$. Moreover, to show how relativistic an ideal Fermi gas is, we introduce the ratio $\zeta=\frac{k_{F}}{m}$.
The system under considering is in the non-relativistic limit if $\zeta\ll 1$, in the ultrarelativistic limit if $\zeta \gg 1$. Since $k_F$ is fixed, $m$ changes as $\zeta$ varies.

\begin{figure}[!ht]
\centering
\includegraphics[width=3.3in]{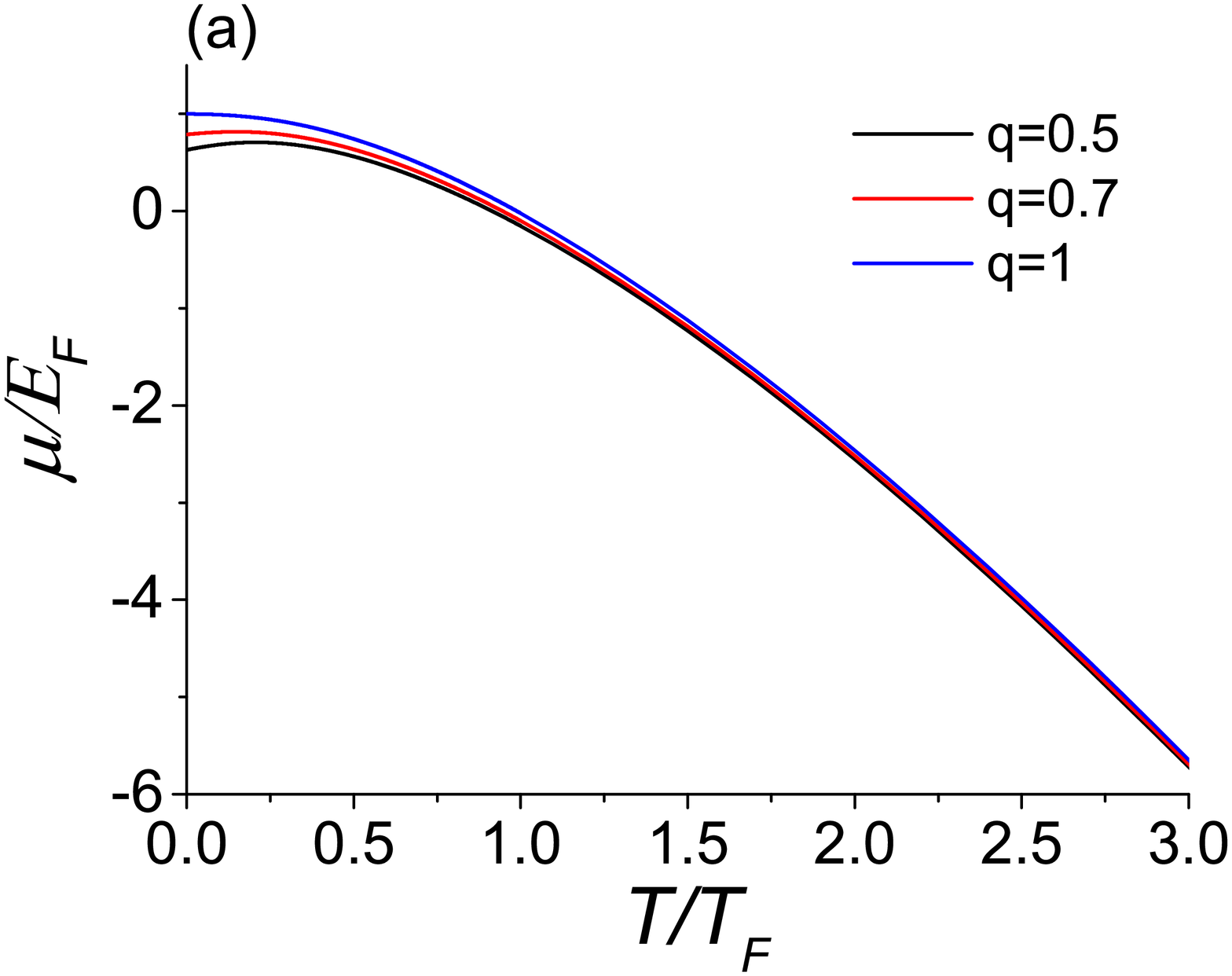}
\includegraphics[width=3.3in]{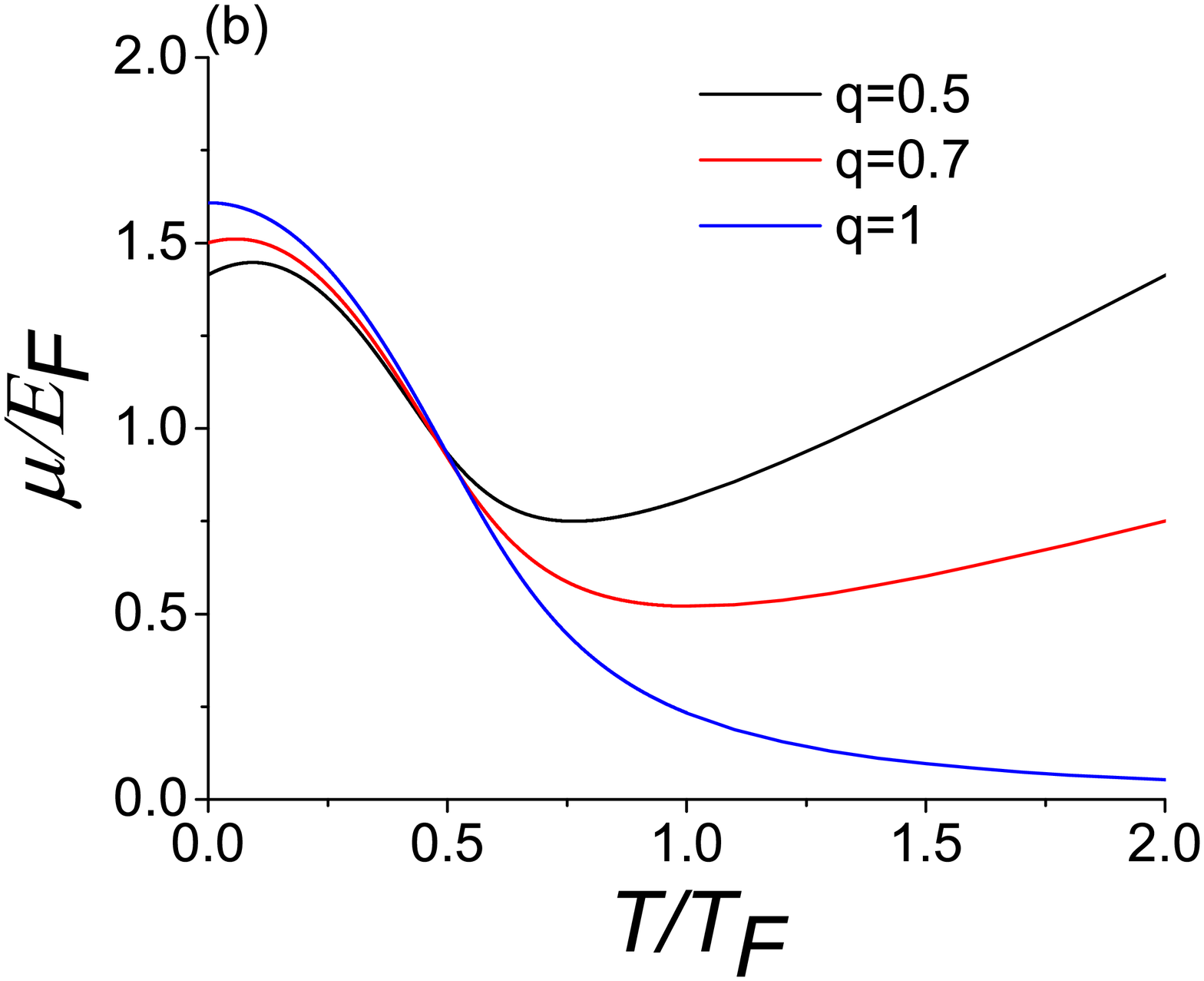}
\caption{(Color online). The chemical potential $\mu$ as a function of the temperature $T$ for (a) $q$-deformed nonrelativistic Fermi gas and (b) $q$-deformed relativistic Fermi gas. The black, red, and blue solid lines correspond to the situation with $q=0.5,0.7,$ and $1.0$ respectively.}
\label{fig.1}
\end{figure}

In Figure.\ref{fig.1}, we plot the chemical potential vs the temperature at several different $q$'s for both $q$-deformed nonrelativistic and relativistic Fermi gases. Panel (a) shows the result for a $q$-deformed nonrelativistic ideal Fermi gas, of which the number equation is  \cite{OurJStat18}
\begin{align}\label{nrn}
n=\sum_\bk\frac{1}{\me^{\beta(\frac{k^2}{2m}-\mu)}+q}.
\end{align}
Panel (b) shows a comparative study of the characteristics of a $q$-deformed relativistic ideal Fermi gas with $\zeta=1.0$. The typical deformation parameters $q=0.5$(black), 0.7(red), and 1.0(blue) are chosen for comparisons.
Here we restrict the values of $q$ in the range $(0,1]$ as pointed out previously.
In the nonrelativistic case, obviously $\mu=E_F$ at $T=0$ for ordinary ideal Fermi gas ($q=1.0$). Moreover, $\mu(T=0)$ is smaller than the ordinary result $E_F$. This is not surprising since $\mu(T=0)=(6q\pi^2n)^{\frac{2}{3}}/(2m)$ where $0<q< 1$ and $n$ is a fixed value. At high temperatures, the chemical potential becomes negative in all three situations according to Fig.\ref{fig.1}(a). Naively, this can be inferred from the fact that the Fermi distribution in Eq.(\ref{nrn}), either undeformed ($q=1.0$) or deformed ($q\neq1.0$), approaches the Maxwell distribution at high temperatures if no relativistic effects are considered. In other words, all quantum effects are smeared by thermal fluctuations, which is independent of what type of quantum algebra is imposed.
However, this does not mean that all high-temperature behaviors of a $q$-deformed nonrelativistic ideal gas are quite similar to its
ordinary counterpart. More complete studies on the thermostatistical properties of a $q$-deformed Fermi gas can be found in Refs.\cite{qaPRE12,qaPA16}.

In Fig.\ref{fig.1}(b) we give the sketch of the results in the moderate relativistic regime with $\zeta=1.0$ as a direct comparison. The low-temperature property is similar to that of Fig.\ref{fig.1}(a) except $\mu(T=0)\neq E_F$ when $q=1.0$. This is because the energy dispersion has changed if a Fermi gas becomes relativistic. However, high-temperature behavior is significantly different.
This does make sense since more antiparticles are excited due to the high energetic thermal fluctuation at higher temperature. Thus, the quantum effect is still dominant at high temperatures.
In the ordinary situation with $q=1.0$, $\mu$ approaches 0 but is still positive. As we have pointed out previously, for a relativistic gas, $\mu-m$ plays the role of nonrelativistic chemical potential $\mu_\text{nr}$ if $|\mathbf{k}|$ is small. Hence, approximately $\mu\approx \mu_\text{nr}+m$, which is always positive.
Moreover, in contrast to Fig.\ref{fig.1}(a) for the nonrelativistic case, $\mu$ starts to increase when temperature is larger than a certain value. The reason is outlined as follows. As we have pointed out, in the deformed situations with $q\neq 1.0$, it is $\mu+T\ln q$ that plays the role of the chemical potential of ordinary Fermi gases. To clarify this more clearly, we define the shifted chemical potential $\mu_s=\mu+T\ln q$, which equals to the ordinary chemical potential at $q=1.0$. If $\mu_s$ behaves like the ordinary chemical potential at $q\neq 1.0$, then the second term of $\mu=\mu_s-T\ln q$, i.e. $-T\ln q$, gives significant effects at high temperatures and makes the asymptotic behavior of $\mu$ the same as in Fig.\ref{fig.1}(b).

\begin{figure}[!ht]
\centering
\includegraphics[width=3.3in]{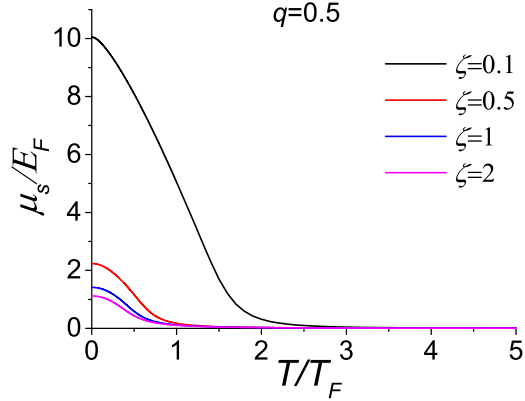}
\caption{(Color online). The shifted chemical potential $\mu_s$ as function of the temperature for $q=0.5$. The black, red, blue and pink solid lines correspond to the situations with $\zeta=0.1,0.5,1$ and $2$ respectively.}
\label{fig.2}
\end{figure}

Now, we need to make sure that $\mu_s$ does behave as expected in the relativistic regimes, and the result is outlined in Figure.\ref{fig.2}. The situations with $q=0.5$ is presented, and $\zeta=0.1,0.5,1$ and $2$ are chosen to denote the regimes from weak to strong relativistic limits. All $\mu_s$'s have a similar shape to the ordinary chemical potential shown in Fig.\ref{fig.1}(b). The only explicit difference is the value of $\mu_s(T=0)$, which depends on the ratio $\zeta$. In the weak relativistic regime with $\zeta=0.1$, the result indicates $\mu_s(T=0)\approx m/\zeta$ since all kinetic degrees of freedom are frozen at $T=0$. Since $\mu=\mu_s-T\ln q=\mu_s+T\ln 2$ and $\mu_s$ behaves like the chemical potential of ordinary fermions (i.e. it asymptotically approaches zero at high $T$, just like the $q=1.0$ line in Fig.\ref{fig.1}(b)), it can be concluded that $\mu$ does have a linearly increasing trend with the slope $-\ln q=\ln 2$ at high temperature as shown in Fig.\ref{fig.1}(b), which is due to the increasing production of $q$-antifermions by the high energetic thermal fluctuation.

\begin{figure}[!ht]
\centering
\includegraphics[width=3.3in]{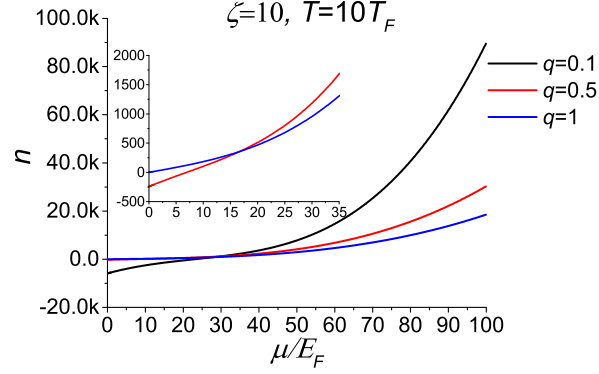}
\caption{(Color online). Particle number as a function of $\mu$ at $\zeta=10$ and $T=10T_F$. The black, red, blue and pink solid lines correspond to the situations with $q=0.1$, 0.5, 1.0 and 2.0 respectively. The inset displays the fine details when $q=0.5$ and 1.0.}
\label{fig.3}
\end{figure}

Since the deformation parameter $q$ has a nontrivial effect on the chemical potential, it will, in turn, produce an implication on the particle number. A quick inspection on Fig.\ref{fig.1}(b) reveals the fact that $\mu$ has a minimum at some finite temperature when $q<1.0$. Thus, it might be easier to create antiparticles within the neighborhood of that point.
Moreover, if a system is in a deeper relativistic regime, it is also easier to produce antiparticles, which gives a negative contribution to the total particle number.
Hence, we consider a more extreme situation with $\zeta=10$ (ultrarelativistic) and $T=10T_F$ (high temperature), and the results are visualized in Figure.\ref{fig.3}. We plot $n$ vs $\mu$ at different $q$'s. As expected, if $q<1.0$, for example $q=0.1$, the total particle number becomes negative in some region of $\mu$ (The inset shows the enlarged structure when $q=0.5$, $n$ also becomes negative when $\mu$ in a certain regime). That is, there are more antiparticles than particles. This is essentially different from ordinary noninteracting Fermi gases. It might be possible in future studies to use a simple theory of $q$-gas to simulate the phase of the very early universe by adjusting the deformation parameter $q$.

\section{Application to White Dwarf}\label{Sec.4b}
As a possible application, we try to use our $q$-deformed model of relativistic Fermi gases to study the electron gas inside a white dwarf, which can be approximated as a nearly-ideal relativistic electron gas. We are interested in the effect of the deformation parameter on Chandrasekhar limit. We consider the simple situation that the white dwarf is made of Helium and electron-degenerate matter. Most electrons stay in the ground state, i.e. the Fermi sphere. In other words, the temperature of the system is much lower than the Fermi temperature. Thus, the contribution of the $q$-deformed antielectrons can be safely ignored according to Eq.(\ref{n0}). Furthermore, Eqs.(\ref{Et}) and (\ref{0TN}) indicates that the ground state energy of degenerate $q$-deformed electron gas is then given by
\begin{align}\label{E0}
E_0=\frac{2}{q}\sum_{|\mathbf{k}|<k_F}\sqrt{\mathbf{k}^2+m^2_e}=\frac{2V}{q(2\pi)^3}\int_0^{k_F}\dif k4\pi k^2\sqrt{\mathbf{k}^2+m^2_e}.
\end{align}
Here the factor $2$ comes from the spin degrees of freedom, and $k_F$ is the Fermi momentum of the $q$-electrons (For simplicity, we still use the symbol $k_F$ to denote the Fermi momentum in this situation, please notice the difference between it and those in previous sections), which is determined by
\begin{align}\label{kFe}
\frac{2V}{q(2\pi)^3}\frac{4\pi}{3}k^3_F=N_e\Rightarrow k_F=\sqrt[3]{3q\pi^2n_e}.
\end{align}
Here $N_e$ is the total number of $q$-electrons, and $n_e=\frac{N_e}{V}$. Let $\zeta=\frac{k}{m_e}$ and $\zeta_F=\frac{k_F}{m_e}$, Eq.(\ref{E0}) implies
\begin{align}\label{E01}
E_0=\frac{m^4_eV}{q\pi^2}f(\zeta_F)
\end{align}
where
\begin{align}\label{fz}
f(\zeta_F)=\int_0^{\zeta_F}\dif \zeta \zeta^2\sqrt{1+\zeta^2}=\frac{1}{4}\zeta_F^4\left(1+\frac{1}{\zeta_F^2}+\cdots\right)
\end{align}
in the relativistic limit.
Note the number of Helium nuclei is $\frac{1}{2}N_e$, then the total mass of the white dwarf is $M=(m_e+2m_p)N_e\approx 2m_pN_e$ where $m_p$ is the mass of proton. Using $V=\frac{4\pi R^3}{3}$ where $R$ is assumed to be the radius of the white dwarf, the following relation can be obtained
\begin{align}\label{tr}
\zeta_F=\frac{1}{m_e R}\sqrt[3]{\frac{9q\pi M}{8m_p}}=\frac{\sqrt[3]{\bar{M}}}{\bar{R}}
\end{align}
where $\bar{M}=\frac{9q\pi M}{8m_p}$, $\bar{R}=m_e R$.

The self gravitational potential energy of the white dwarf is $E_G=-\frac{3}{5}\frac{GM^2}{R}$ where $G$ is the gravitational constant, then the total energy is $E_0+E_G$. When a white dwarf is stable, its total energy must be at the minimum, which requires
\begin{align}\label{ec}
\frac{\partial (E_0+E_G)}{\partial R}=0\Rightarrow P_04\pi R^2-\frac{3}{5}\frac{GM^2}{R^2}=0, \quad \text{or}\quad P_0=\frac{3}{20\pi}\frac{GM^2}{R^4}=\frac{3m_e^4G}{20\pi}\left(\frac{8m_p}{9q\pi}\right)^2\frac{\bar{M}^2}{\bar{R}^4}.
\end{align}
Here $P_0$ is the degeneracy pressure of the $q$-electron gas, given by
\begin{align}\label{pwd}
P_0=-\frac{\partial E_0}{\partial V}=\frac{m^4_e}{q\pi^2}\left(\frac{1}{3}\zeta^3_F\sqrt{1+\zeta_F^2}-f(\zeta_F)\right)\approx \frac{m^4_e}{12q\pi^2}(\zeta_F^4-\zeta_F^2)=\frac{m^4_e}{12q\pi^2}\left(\frac{\bar{M}^{\frac{4}{3}}}{\bar{R}^4}-\frac{\bar{M}^{\frac{2}{3}}}{\bar{R}^2}\right)
\end{align}
in the relativistic limit, where Eq.(\ref{tr}) has been plugged in. Solving Eqs.(\ref{ec}) and (\ref{pwd}), we get
\begin{align}\label{fe}
\bar{R}=\bar{M}^{\frac{1}{3}}\sqrt{1-\left(\frac{\bar{M}}{\bar{M}_0}\right)^{\frac{2}{3}}}
\end{align}
where
 \begin{align}\label{M0}
\bar{M}_0=\left(\frac{45q\pi}{64Gm_p^2}\right)^{\frac{3}{2}}.
\end{align}
Thus, the condition $\bar{R}>0$ leads to the critical mass of the white dwarf
 \begin{align}\label{M1}
M_0(q)=\frac{8m_p}{9q\pi}\bar{M}_0=\sqrt{q}\frac{8}{9\pi}\left(\frac{45\pi}{64}\right)^{\frac{3}{2}}\left(\frac{\hbar c}{G m^2_p}\right)^{\frac{3}{2}}m_p=\sqrt{q}M_0(q=1)\approx \sqrt{q}\times 1.7 M_{\bigodot},
\end{align}
where we have restored the dependence on $\hbar$ and $c$, $M_0(q=1)$ is the ordinary critical mass of white dwarfs, and $M_{\bigodot}$ is the mass of the sun.
 Interestingly, the critical mass has a simple dependence $\sqrt{q}$ on the deformation parameter.
 If we include more high order terms in the expansion of the integral (\ref{fz}), we can get a more accurate estimation of $M_0(q=1)$ and also a more complicated dependence on $q$ of the critical mass $M_0(q)$.

\section{Conclusion}\label{Sec.5}

In summary, we construct a self-consistent $q$-deformed algebra for a quantum system containing both particles and antiparticles. Based on this algebra, we further formulate the finite-temperature many-body theory for the $q$-deformed relativistic ideal Fermi gas. By applying this formalism, we study the thermodynamic properties of this system. Interestingly, the $q$-deformed relativistic Fermi gas presents significantly different features from ordinary Fermi gases even in the noninteracting scenario. The deformation parameter has a notable effect on the chemical potential of both particle and antiparticle excitations, which induces the reversion of particle/antiparticle numbers when $q$ is very small in the ultrarelativistic limit. These findings will lay a solid ground for future studies on the $q$-deformed relativistic interacting Fermi gas. As an application, we study the electron gas in a white dwarf based on our model of relativistic $q$-gas.

Hao Guo thanks the support from the National Natural Science
Foundation of China (Grant No. 11674051), and Xu-Yang Hou thanks the support from the Scientific Research Foundation of Graduate School of Southeast University(Grant No. YBPY2029).

\appendix

\section{Fermionic basic numbers}\label{app0}
To find the fermionic basic numbers $[n_\bk]_\mp$, we assume that the actions of the deformed fermion operators on the Fock states are
\begin{align}\label{aa}
\frac{a_{\bk}}{\sqrt{(2\pi)^3}}|n_\bk\rangle_-=C_{n_\bk}|n_\bk-1\rangle_-,\quad \frac{a^\dagger_{\bk}}{\sqrt{(2\pi)^3}}|n_\bk\rangle_-=C'_{n_\bk}|n_\bk+1\rangle_-
\end{align}
It can be shown $[\hat{n}^-_\bk]|n_\bk\rangle_-=C_{n_\bk}C'_{n_\bk-1}|n_\bk\rangle_-$, i.e. $[n_\bk]_-=C_{n_\bk}C'_{n_\bk-1}$. Similarly, we have
\begin{align}\label{aa2}
\frac{1}{(2\pi)^3}a_{\bk}a^\dagger_{\bk}|n_\bk\rangle_-=C_{n_\bk+1}C'_{n_\bk}|n_\bk\rangle_-=[n_\bk+1]_-|n_\bk\rangle_-.
\end{align}
Applying these relations to the first identity in Eqs.(\ref{a1}), we get
\begin{align}\label{aa3}
[n_\bk+1]_-=1-q[n_\bk]_-.
\end{align}
Taking the choice $[0]_-=0$, this recurrence equation is solved as
\begin{align}\label{aa4}
[n_\bk]_-=\frac{1-(-1)^{n_{\bk}}q^{n_{\bk}}}{1+q}
\end{align}
which is the fermionic basic number of the $q$-fermions. The above analysis also implies $C_{n_\bk}=\sqrt{[n_\bk]_-}$ and $C'_{n_\bk}=\sqrt{[n_\bk+1]_-}$. Thus, the Fock space of $q$-fermions can be constructed on the ground state successively as
\begin{align}\label{aa5}
|n_\bk\rangle_-=\frac{1}{\sqrt{(2\pi)^{3n_\bk}}}\frac{(a^\dagger_\bk)^{n_\bk}}{\sqrt{[n_\bk]_-!}}|0\rangle
\end{align}
where $[n_\bk]_-!=[n_\bk]_-[n_\bk-1]_-\cdots [2]_-[1]_-$.

\section{Conventions of Notations}\label{app1}
In this paper we adopt the Weyl or chiral representation of the gamma matrices,

\begin{equation}
\gamma^{0}=
\begin{pmatrix}\displaystyle
0 & I \\
I & 0
\end{pmatrix}
,\mbox{ } \gamma^{i}=-\gamma_{i}=
\begin{pmatrix}\displaystyle
0 & \sigma^{i} \\
-\sigma^{i} & 0
\end{pmatrix}
,\mbox{ } \gamma_{5}=
\begin{pmatrix}\displaystyle
-I & 0 \\
0 & I
\end{pmatrix}.
\end{equation}
Define the 4-vector pauli matrices as
\begin{equation}
\sigma^{\mu}=(1,\vec{\sigma}),\qquad \bar{\sigma}^{\mu}=(1,-\vec{\sigma}),
\end{equation}
then
\begin{equation}
\gamma^\mu=\begin{pmatrix}\displaystyle
0 & \sigma^\mu \\
\bar{\sigma}^\mu & 0
\end{pmatrix}.
\end{equation}
The scalar between two 4-vectors, for example, $k^\mu$ and $\sigma^\mu$, is defined as $k\cdot\sigma=\epsilon_\mathbf{k}-\mathbf{k}\cdot\vec{\sigma}$ , while $\mi  k\cdot x=\mi\epsilon_\bk t-\mi\mathbf{k}\cdot\mathbf{x}=\epsilon_\bk\tau-\mi\mathbf{k}\cdot\mathbf{x}$.

\section{$q$-deformed Anti-commutative Relation}\label{app2}
Note the time-independent fermion field can be expanded as
\begin{align}\label{psi}
\psi (\mathbf{x})&=\sum_{\mathbf{k}}\frac{1}{\sqrt{\epsilon_{\mathbf{k}}}}\left\{
\left[ \begin{array}{ccc}
\sqrt{k\cdot\sigma}\eta^{L}_{\mathbf{k}} \\
\sqrt{k\cdot\bar{\sigma}}\eta^{L}_{\mathbf{k}}
\end{array}\right]a_\mathbf{k}\me^{\mi \mathbf{k}\cdot \mathbf{x}}+
\left[ \begin{array}{ccc}
\sqrt{k\cdot\sigma}\eta^{R}_{\mathbf{k}} \\
-\sqrt{k\cdot\bar{\sigma}}\eta^{R}_{\mathbf{k}}
\end{array}\right]b^{\dagger}_\mathbf{k}\me^{-\mi \mathbf{k}\cdot \mathbf{x}}\right\},\nonumber\\
\psi^\dagger (\mathbf{x})&=\sum_{\mathbf{k}}\frac{1}{\sqrt{\epsilon_{\mathbf{k}}}}\left\{
\left[ \begin{array}{l}
\eta^{L\dagger}_{\mathbf{k}}\sqrt{k\cdot\sigma} ,
\eta^{L\dagger}_{\mathbf{k}}\sqrt{k\cdot\bar{\sigma}}
\end{array}\right]a^\dagger_\mathbf{k}\me^{-\mi \mathbf{k}\cdot  \mathbf{x}}+
\left[ \begin{array}{l}
\eta^{R\dagger}_{\mathbf{k}} \sqrt{k\cdot\sigma},
-\eta^{R\dagger}_{\mathbf{k}}\sqrt{k\cdot\bar{\sigma}}
\end{array}\right]b_\mathbf{k}\me^{\mi \mathbf{k}\cdot  \mathbf{x}}\right\}.
\end{align}
By applying these expressions and the algebras (\ref{a1}), it can be shown that
\begin{align}\label{t1}
&\psi_a (\mathbf{x})\psi_b^\dagger (\mathbf{x}')+q\psi_b^\dagger (\mathbf{x}')\psi_a (\mathbf{x})\notag\\
&=\sum_{\mathbf{k}\mathbf{k}'}\frac{1}{\sqrt{\epsilon_{\mathbf{k}}\epsilon_{\mathbf{k}'}}}\bigg\{
\begin{pmatrix}\displaystyle
\sqrt{k\cdot\sigma}\eta^{L}_{\mathbf{k}}\eta^{L\dagger}_{\mathbf{k}'}\sqrt{k'\cdot\sigma}  & \sqrt{k\cdot\sigma}\eta^{L}_{\mathbf{k}}\eta^{L\dagger}_{\mathbf{k}'}\sqrt{k'\cdot\bar{\sigma}}  \\
\sqrt{k\cdot\bar{\sigma}}\eta^{L}_{\mathbf{k}}\eta^{L\dagger}_{\mathbf{k}'}\sqrt{k'\cdot\sigma} &
\sqrt{k\cdot\bar{\sigma}}\eta^{L}_{\mathbf{k}}\eta^{L\dagger}_{\mathbf{k}'}\sqrt{k'\cdot\bar{\sigma}}
\end{pmatrix}_{ab}(a_{\bk}a^\dagger_{\bk'}+qa^\dagger_{\bk'}a_{\bk})\me^{\mi \mathbf{k}\cdot \mathbf{x}-\mi \mathbf{k}'\cdot \mathbf{x}'}\notag\\
&\quad\quad\quad\quad\quad+\begin{pmatrix}\displaystyle
\sqrt{k\cdot\sigma}\eta^{L}_{\mathbf{k}}\eta^{R\dagger}_{\mathbf{k}'}\sqrt{k'\cdot\sigma}  & -\sqrt{k\cdot\sigma}\eta^{L}_{\mathbf{k}}\eta^{R\dagger}_{\mathbf{k}'}\sqrt{k'\cdot\bar{\sigma}}  \\
\sqrt{k\cdot\bar{\sigma}}\eta^{L}_{\mathbf{k}}\eta^{R\dagger}_{\mathbf{k}'}\sqrt{k'\cdot\sigma} &
-\sqrt{k\cdot\bar{\sigma}}\eta^{L}_{\mathbf{k}}\eta^{R\dagger}_{\mathbf{k}'}\sqrt{k'\cdot\bar{\sigma}}
\end{pmatrix}_{ab}(a_{\bk}b_{\bk'}+qb_{\bk'}a_{\bk})\me^{\mi \mathbf{k}\cdot \mathbf{x}+\mi \mathbf{k}'\cdot \mathbf{x}'}\notag\\
&\quad\quad\quad\quad\quad+\begin{pmatrix}\displaystyle
\sqrt{k\cdot\sigma}\eta^{R}_{\mathbf{k}}\eta^{L\dagger}_{\mathbf{k}'}\sqrt{k'\cdot\sigma}  & \sqrt{k\cdot\sigma}\eta^{R}_{\mathbf{k}}\eta^{L\dagger}_{\mathbf{k}'}\sqrt{k'\cdot\bar{\sigma}}  \\
-\sqrt{k\cdot\bar{\sigma}}\eta^{R}_{\mathbf{k}}\eta^{L\dagger}_{\mathbf{k}'}\sqrt{k'\cdot\sigma} &
-\sqrt{k\cdot\bar{\sigma}}\eta^{R}_{\mathbf{k}}\eta^{L\dagger}_{\mathbf{k}'}\sqrt{k'\cdot\bar{\sigma}}
\end{pmatrix}_{ab}(b^\dagger_{\bk}a^\dagger_{\bk'}+qa^\dagger_{\bk'}b^\dagger_{\bk})\me^{-\mi \mathbf{k}\cdot \mathbf{x}-\mi \mathbf{k}'\cdot \mathbf{x}'}\notag\\
&\quad\quad\quad\quad\quad+\begin{pmatrix}\displaystyle
\sqrt{k\cdot\sigma}\eta^{R}_{\mathbf{k}}\eta^{R\dagger}_{\mathbf{k}'}\sqrt{k'\cdot\sigma}  & -\sqrt{k\cdot\sigma}\eta^{R}_{\mathbf{k}}\eta^{R\dagger}_{\mathbf{k}'}\sqrt{k'\cdot\bar{\sigma}}  \\
-\sqrt{k\cdot\bar{\sigma}}\eta^{R}_{\mathbf{k}}\eta^{R\dagger}_{\mathbf{k}'}\sqrt{k'\cdot\sigma} &
\sqrt{k\cdot\bar{\sigma}}\eta^{R}_{\mathbf{k}}\eta^{R\dagger}_{\mathbf{k}'}\sqrt{k'\cdot\bar{\sigma}}
\end{pmatrix}_{ab}
(b^\dagger_{\bk}b_{\bk'}+qb_{\bk'}b^\dagger_{\bk})\me^{-\mi \mathbf{k}\cdot \mathbf{x}+\mi \mathbf{k}'\cdot \mathbf{x}'}\bigg\}\notag\\
&=\sum_{\mathbf{k}}\frac{(2\pi)^3}{\epsilon_\bk}\bigg\{
\begin{pmatrix}\displaystyle
(\epsilon_\bk+|\mathbf{k}|)\eta^{L}_{\mathbf{k}}\eta^{L\dagger}_{\mathbf{k}} &
m\eta^{L}_{\mathbf{k}}\eta^{L\dagger}_{\mathbf{k}} \\
m\eta^{L}_{\mathbf{k}}\eta^{L\dagger}_{\mathbf{k}} &
(\epsilon_\bk-|\mathbf{k}|)\eta^{L}_{\mathbf{k}}\eta^{L\dagger}_{\mathbf{k}}
\end{pmatrix}_{ab}\me^{\mi \mathbf{k}\cdot (\mathbf{x}-\mathbf{x}')}
+\begin{pmatrix}\displaystyle
(\epsilon_\bk-|\mathbf{k}|)\eta^{R}_{\mathbf{k}}\eta^{R\dagger}_{\mathbf{k}} &
-m\eta^{R}_{\mathbf{k}}\eta^{R\dagger}_{\mathbf{k}} \\
-m\eta^{R}_{\mathbf{k}}\eta^{R\dagger}_{\mathbf{k}} &
(\epsilon_\bk+|\mathbf{k}|)\eta^{R}_{\mathbf{k}}\eta^{R\dagger}_{\mathbf{k}}
\end{pmatrix}_{ab}\me^{-\mi \mathbf{k}\cdot (\mathbf{x}-\mathbf{x}')}\bigg\},
\end{align}
where we have applied the fact $(k\cdot\sigma)\eta^{L}_{\mathbf{k}}=(\epsilon_\bk+|\mathbf{k}|)\eta^{L}_{\mathbf{k}}$ and $(k\cdot\sigma)\eta^{R}_{\mathbf{k}}=(\epsilon_\bk-|\mathbf{k}|)\eta^{R}_{\mathbf{k}}$.
When the vector changes as $\mathbf{k}\rightarrow-\mathbf{k}$, the polar and azimuthal angles satisfy $\theta_{-\mathbf{k}}=\pi-\theta_{\mathbf{k}}$ and $\phi_{-\mathbf{k}}=\pi+\phi_{\mathbf{k}}$. Note
\begin{align}
\eta^{L}_{\mathbf{k}}\eta^{L\dagger}_{\mathbf{k}}=\begin{pmatrix}\displaystyle
\sin^2\frac{\theta_\bk}{2} &
-\sin\frac{\theta_\bk}{2}\cos\frac{\theta_\bk}{2}\me^{-\mi\phi_\bk} \\
-\sin\frac{\theta_\bk}{2}\cos\frac{\theta_\bk}{2}\me^{\mi\phi_\bk} &
\cos^2\frac{\theta_\bk}{2}
\end{pmatrix},\quad
\eta^{R}_{\mathbf{k}}\eta^{R\dagger}_{\mathbf{k}}=\begin{pmatrix}\displaystyle
\cos^2\frac{\theta_\bk}{2} &
\sin\frac{\theta_\bk}{2}\cos\frac{\theta_\bk}{2}\me^{-\mi\phi_\bk} \\
\sin\frac{\theta_\bk}{2}\cos\frac{\theta_\bk}{2}\me^{\mi\phi_\bk} &
\sin^2\frac{\theta_\bk}{2}
\end{pmatrix}
\end{align}
Hence we have $\eta^{L}_{-\mathbf{k}}\eta^{L\dagger}_{-\mathbf{k}}=\eta^{R}_{\mathbf{k}}\eta^{R\dagger}_{\mathbf{k}}$ and $\eta^{L}_{\mathbf{k}}\eta^{L\dagger}_{\mathbf{k}}+\eta^{R}_{\mathbf{k}}\eta^{R\dagger}_{\mathbf{k}}=\mathbf{1}_{2\times2}$. Therefore, by changing the variable as $\mathbf{k}\rightarrow-\mathbf{k}$ in the second term of Eq.(\ref{t1}), we have
\begin{align}\label{t2}
\psi_a (\mathbf{x})\psi_b^\dagger (\mathbf{x}')+q\psi_b^\dagger (\mathbf{x}')\psi_a (\mathbf{x})=2(2\pi)^3\sum_{\mathbf{k}}
\begin{pmatrix}\displaystyle
\eta^{L}_{\mathbf{k}}\eta^{L\dagger}_{\mathbf{k}} &
0 \\
0 &
\eta^{L}_{\mathbf{k}}\eta^{L\dagger}_{\mathbf{k}}
\end{pmatrix}_{ab}\me^{\mi \mathbf{k}\cdot (\mathbf{x}-\mathbf{x}')}=2(2\pi)^3\sum_{\mathbf{k}}(\mathbf{1}_{2\times2}\otimes\eta^{L}_{\mathbf{k}}\eta^{L\dagger}_{\mathbf{k}})_{ab}\me^{\mi \mathbf{k}\cdot (\mathbf{x}-\mathbf{x}')}.
\end{align}
Similarly
\begin{align}\label{t3}
\psi_a (\mathbf{x})\psi_b^\dagger (\mathbf{x}')+q\psi_b^\dagger (\mathbf{x}')\psi_a (\mathbf{x})=2(2\pi)^3\sum_{\mathbf{k}}
\begin{pmatrix}\displaystyle
\eta^{R}_{\mathbf{k}}\eta^{R\dagger}_{\mathbf{k}} &
0 \\
0 &
\eta^{R}_{\mathbf{k}}\eta^{R\dagger}_{\mathbf{k}}
\end{pmatrix}_{ab}\me^{-\mi \mathbf{k}\cdot (\mathbf{x}-\mathbf{x}')}=2(2\pi)^3\sum_{\mathbf{k}}(\mathbf{1}_{2\times2}\otimes\eta^{R}_{\mathbf{k}}\eta^{R\dagger}_{\mathbf{k}})_{ab}\me^{-\mi \mathbf{k}\cdot (\mathbf{x}-\mathbf{x}')}.
\end{align}
Then evaluate ((\ref{t2})+(\ref{t3}))/2, we get
\begin{align}\label{t4}
\psi_a (\mathbf{x})\psi_b^\dagger (\mathbf{x}')+q\psi_b^\dagger (\mathbf{x}')\psi_a (\mathbf{x})&=(2\pi)^3\sum_{\mathbf{k}}\Big\{(\mathbf{1}_{2\times2}\otimes\mathbf{1}_{2\times2})_{ab}\cos\left[\mathbf{k}\cdot (\mathbf{x}- \mathbf{x}')\right]
+\mi\left[\mathbf{1}_{2\times2}\otimes(\eta^{L}_{\mathbf{k}}\eta^{L\dagger}_{\mathbf{k}}-\eta^{R}_{\mathbf{k}}\eta^{R\dagger}_{\mathbf{k}})\right]_{ab}\sin\left[\mathbf{k}\cdot (\mathbf{x}-\mathbf{x}')\right]\Big\}\notag\\
&=(2\pi)^3\sum_{\mathbf{k}}\delta_{ab}\frac{\me^{\mi \mathbf{k}\cdot (\mathbf{x}-\mathbf{x}')}+\me^{-\mi \mathbf{k}\cdot (\mathbf{x}-\mathbf{x}')}}{2}\notag\\
&=\delta_{ab}\delta(\mathbf{x}-\mathbf{x}'),
\end{align}
where in the second line we have applied the fact that every element of $\eta^{L}_{\mathbf{k}}\eta^{L\dagger}_{\mathbf{k}}-\eta^{R}_{\mathbf{k}}\eta^{R\dagger}_{\mathbf{k}}$ contains $\me^{\pm\mi\phi_\bk}$ and $\int_0^{2\pi}\dif\phi_\bk\me^{\pm\mi\phi_\bk}=0$.

\section{Equation of Motion of the Fermion Field}\label{app3}
Applying the Heisenberg equation and the algebras (\ref{a1}), we have
\begin{align}
&\gamma^0\frac{\partial \psi(x)}{\partial \tau}-(\mi\vec{\gamma}\cdot\nabla-m+\mu\gamma^0) \psi(x)\notag\\
&=\sum_{\mathbf{k}}\frac{1}{\sqrt{\epsilon_{\mathbf{k}}}}\gamma^0\Big\{
-\left[ \begin{array}{ccc}
\sqrt{k\cdot\sigma}\eta^{L}_{\mathbf{k}} \\
\sqrt{k\cdot\bar{\sigma}}\eta^{L}_{\mathbf{k}}
\end{array}\right](\epsilon_\bk-\mu)a_\mathbf{k}\me^{-\mi k_-\cdot x}+
\left[ \begin{array}{ccc}
\sqrt{k\cdot\sigma}\eta^{R}_{\mathbf{k}} \\
-\sqrt{k\cdot\bar{\sigma}}\eta^{R}_{\mathbf{k}}
\end{array}\right](\epsilon_\bk+\mu)b^{\dagger}_\mathbf{k}\me^{\mi k_+\cdot x}\Big\}\notag\\
&-\sum_{\mathbf{k}}\frac{1}{\sqrt{\epsilon_{\mathbf{k}}}}\Big\{
(-\vec{\gamma}\cdot\mathbf{k}-m+\mu\gamma^0)\left[ \begin{array}{ccc}
\sqrt{k\cdot\sigma}\eta^{L}_{\mathbf{k}} \\
\sqrt{k\cdot\bar{\sigma}}\eta^{L}_{\mathbf{k}}
\end{array}\right]a_\mathbf{k}\me^{-\mi k_-\cdot x}+
(\vec{\gamma}\cdot\mathbf{k}-m+\mu\gamma^0)\left[ \begin{array}{ccc}
\sqrt{k\cdot\sigma}\eta^{R}_{\mathbf{k}} \\
-\sqrt{k\cdot\bar{\sigma}}\eta^{R}_{\mathbf{k}}
\end{array}\right]b^{\dagger}_\mathbf{k}\me^{\mi k_+\cdot x}\Big\}\notag\\
&=\sum_{\mathbf{k}}\frac{1}{\sqrt{\epsilon_{\mathbf{k}}}}\Big\{
\begin{pmatrix}\displaystyle
m & -k\cdot\sigma \\
-k\cdot\bar{\sigma} & m
\end{pmatrix}\left[ \begin{array}{ccc}
\sqrt{k\cdot\sigma}\eta^{L}_{\mathbf{k}} \\
\sqrt{k\cdot\bar{\sigma}}\eta^{L}_{\mathbf{k}}
\end{array}\right]a_\mathbf{k}\me^{-\mi k_-\cdot x}
+
\begin{pmatrix}\displaystyle
m & k\cdot\sigma \\
k\cdot\bar{\sigma} & m
\end{pmatrix}
\left[ \begin{array}{ccc}
\sqrt{k\cdot\sigma}\eta^{R}_{\mathbf{k}} \\
-\sqrt{k\cdot\bar{\sigma}}\eta^{R}_{\mathbf{k}}
\end{array}\right]b^{\dagger}_\mathbf{k}\me^{\mi k_+\cdot x}\Big\}\notag\\
&=\sum_{\mathbf{k}}\frac{1}{\sqrt{\epsilon_{\mathbf{k}}}}\Big\{
\left[ \begin{array}{ccc}
(m\sqrt{k\cdot\sigma}-k\cdot\sigma\sqrt{k\cdot\bar{\sigma}})\eta^{L}_{\mathbf{k}} \\
(-k\cdot\bar{\sigma}\sqrt{k\cdot\sigma}+m\sqrt{k\cdot\bar{\sigma}})\eta^{L}_{\mathbf{k}}
\end{array}\right]a_\mathbf{k}\me^{-\mi k_-\cdot x}+
\left[ \begin{array}{ccc}
(m\sqrt{k\cdot\sigma}-k\cdot\sigma\sqrt{k\cdot\bar{\sigma}})\eta^{R}_{\mathbf{k}} \\
(k\cdot\bar{\sigma}\sqrt{k\cdot\sigma}-m\sqrt{k\cdot\bar{\sigma}})\eta^{R}_{\mathbf{k}}
\end{array}\right]b^{\dagger}_\mathbf{k}\me^{\mi k_+\cdot x}\Big\}\notag\\
&=0,
\end{align}
where we have applied the fact $(k\cdot\sigma)(k\cdot\bar{\sigma})=(k\cdot\bar{\sigma})(k\cdot\sigma)=m^2$. This leads to the equation of motion (\ref{eom}) for $q$-deformed fermionic field.

\section{Expression of the Green's Function}\label{app4}
It is easy to verify the following properties of the energy projectors
\begin{equation}
\Lambda^{2}_{\pm}(\mathbf{k})=\Lambda_{\pm}(\mathbf{k}),\mbox{  }\Lambda_{+}(\mathbf{k})\Lambda_{-}(\mathbf{k})=\Lambda_{-}(\mathbf{k})\Lambda_{+}(\mathbf{k})=0
\end{equation}
\begin{equation}
\Lambda_{+}(\mathbf{k})+\Lambda_{-}(\mathbf{k})=\mathbf{1}_{4\times4},\mbox{  }\Lambda_{+}(\mathbf{k})-\Lambda_{-}(\mathbf{k})=\frac{1}{\epsilon_{\mathbf{k}}}\gamma^{0}(\vec{\gamma}\cdot\mathbf{k}+m).
\end{equation}
The gamma matrice satisfy
\begin{equation}
(a+b\gamma^{0}+c\vec{\gamma}\cdot\mathbf{k}+d\gamma^{0}\vec{\gamma}\cdot\mathbf{k})^{-1}=\frac{-a+b\gamma^{0}+c\vec{\gamma}\cdot\mathbf{k}+d\gamma^{0}\vec{\gamma}\cdot\mathbf{k}}{b^{2}-a^{2}+(d^{2}-c^{2})|\mathbf{k}|^{2}}.
\end{equation}
By using this equality we have
\begin{eqnarray}
G(K)&=&[(\mi\omega_{n}+\frac{\ln q}{\beta}+\mu)\gamma^{0}-\vec{\gamma}\cdot\mathbf{k}-m]^{-1} \nonumber \\
&=&\frac{(\mi\omega_{n}+\frac{\ln q}{\beta}+\mu )\gamma^{0}-\vec{\gamma}\cdot\mathbf{k}+m}{(\mi\omega_{n}+\frac{\ln q}{\beta}+\mu )^{2}-m^{2}-|\mathbf{k}|^{2}} \nonumber \\
&=&\frac{[(\mi\omega_{n}+\frac{\ln q}{\beta}+\xi^{+}_{\mathbf{k}})\Lambda_{+}(\mathbf{k})+(\mi\omega_{n}+\frac{\ln q}{\beta}-\xi^{-}_{\mathbf{k}})\Lambda_{-}(\mathbf{k})]\gamma^{0}}{(\mi\omega_{n}+\frac{\ln q}{\beta}+\xi^{+}_{\mathbf{k}})(\mi\omega_{n}+\frac{\ln q}{\beta}-\xi^{-}_{\mathbf{k}})} \nonumber\\
&=&\Big[\frac{\Lambda_{+}(\mathbf{k})}{\mi\omega_{n}+\frac{\ln q}{\beta}-\xi^{-}_{\mathbf{k}}}+\frac{\Lambda_{-}(\mathbf{k})}{\mi\omega_{n}+\frac{\ln q}{\beta}+\xi^{+}_{\mathbf{k}}}\Big]\gamma^{0}
\end{eqnarray}

\section{Number equation}\label{app5}
The number equation given by Eq.(\ref{n2}) can be directly verified as follows
\begin{align}
\frac{1}{q}z\frac{\partial \ln Z}{\partial z}&=\frac{1}{q}\sum_\bk\left(\frac{qz\me^{-\beta\epsilon_\bk}}{1+qz\me^{-\beta\epsilon_\bk}}-\frac{1}{z^2}
\frac{q^{-1}z\me^{-\beta\epsilon_\bk}}{1+q^{-1}z^{-1}\me^{-\beta\epsilon_\bk}}\right)\notag\\
&=\sum_\bk\left(\frac{1}{z^{-1}\me^{\beta\epsilon_\bk}+q}-\frac{1}{q}
\frac{1}{1+qz\me^{\beta\epsilon_\bk}}\right)\notag\\
&=\sum_\bk\left(\frac{1}{\me^{\beta(\epsilon_\bk-\mu)}+q}-\frac{1}{q}
\frac{1}{q\me^{\beta(\epsilon_\bk+\mu)}+1}\right)\notag\\
&=n.
\end{align}

\end{document}